\documentclass[%
 reprint,
superscriptaddress,
 amsmath,amssymb,
 aps,prl,
]{revtex4-1}

\usepackage{todonotes}
\usepackage{graphicx}%
\usepackage{dcolumn}%
\usepackage{bm}%
\usepackage[normalem]{ulem}
\usepackage{microtype}
\usepackage{xcolor}
\usepackage{amsmath,amssymb,amsfonts,mathtools,cancel}
\usepackage{multirow}

\definecolor{DragonGreen}{RGB}{0,126,48}

\definecolor{Brownie}{RGB}{97,16,9}

\usepackage[paperwidth=210mm,
            paperheight=297mm,
            left=20mm,
            top=10mm,
            textwidth=170mm,
            marginparsep=3mm,
            marginparwidth=30mm,
            textheight=730pt,
            footskip=50pt]
           {geometry}

\newcommand{\avg}[1]{\ensuremath{\langle{#1}\rangle}}

\newcommand{\gsl}{\ensuremath{\gamma_{\text{sl}}}}
\newcommand{\gp}{\ensuremath{\gamma_{\infty}}}
\newcommand{\ns}{\ensuremath{{n_{\text{s}}}}} %

\newcommand{\ncut}{\ensuremath{n_{\text{cut}}}}
\newcommand{\vs}{\ensuremath{v_{\text{s}}}}

\newcommand{\ms}{\ensuremath{\mu_{\text{s}}}} %
\newcommand{\ml}{\ensuremath{\mu_{\text{l}}}}
\newcommand{\msl}{\ensuremath{\mu_{\text{sl}}}} %

\newcommand{\cond}[2]{\ensuremath{\left(#1\middle|#2\right)}}
\newcommand{\D}{\ensuremath{\mathrm{d}}}

\newcommand{\rhosl}{\ensuremath{\rho_\mathrm{sl}}}

\begin{document}

\preprint{APS/123-QED}

\title{
The Gibbs free energy of homogeneous nucleation:
from atomistic nuclei to the planar limit
}%

\author{Bingqing Cheng}
\email{bingqing.cheng@epfl.ch}
 \affiliation{Laboratory of Computational Science and Modeling, Institute of Materials, {\'E}cole Polytechnique F{\'e}d{\'e}rale de Lausanne, 1015 Lausanne, Switzerland}%

\author{Gareth A. Tribello}
\affiliation{Atomistic Simulation Centre, School of Mathematics and Physics, Queen's 
University Belfast, Belfast, BT7 1NN
}%

\author{Michele Ceriotti}
\affiliation{Laboratory of Computational Science and Modeling, Institute of Materials, {\'E}cole Polytechnique F{\'e}d{\'e}rale de Lausanne, 1015 Lausanne, Switzerland}%

\date{\today}%

\begin{abstract}
In this paper we discuss how the information contained in atomistic simulations of homogeneous nucleation should be used when fitting the parameters in macroscopic nucleation models.  We show how the number of solid and liquid atoms in such simulations can be determined unambiguously by using a Gibbs dividing surface and how the free energy as a function of the number of solid atoms in the nucleus can thus be extracted.  
We then show that the parameters of a model based on classical nucleation theory can be fit using the information contained in these free-energy profiles but that the parameters in such models are highly correlated.  This correlation is unfortunate as it ensures that small errors in the computed free energy surface can give rise to large errors in the extrapolated properties of the fitted model.  To resolve this problem we thus propose a method for fitting macroscopic nucleation models that uses simulations of planar interfaces and simulations of three-dimensional nuclei in tandem.  We show that when the parameters of the macroscopic model are fitted in this way the numerical errors for the final fitted model are smaller and that the extrapolated predictions for large nuclei are thus more reliable. 
\end{abstract}
\maketitle

\section{Introduction}

Nucleation is an important component of many technological and natural processes, 
including metal casting, the assembly of microtubules in cells and the formation of water droplets and ice crystals in the atmosphere
~\cite{oxtoby1992homogeneous,yi2012molecular,sosso2016crystal}.
In spite of its generic importance, however,
our understanding of nucleation kinetics is somewhat unsatisfactory, even for the simplest cases of homogeneous nucleation. %
The main reason for this is that the experimental determination of nucleation rates is challenging. Many of the available analytical techniques can only be used to perform ex situ analysis, which cannot fully capture the dynamical processes that take place during nucleation.

A considerable number of studies in the last two decades have thus used atomistic simulation to study homogeneous nucleation~\cite{ten1996numerical,ten1999homogeneous,moroni2005interplay,trudu2006freezing,lechner2011role,prestipino2012systematic,yi2012molecular,mccarty2016bespoke}. 
Much of this computational effort has concentrated on investigating how the free energy changes with cluster size $G(n)$ because the homogeneous nucleation barrier $G^{\star}$ is the maximum of $G(n)$ and because the nucleation rate can be determined from $G^{\star}$ by using transition state theory or the Bennett-Chandler approach~\cite{ten1996numerical,moroni2005interplay,jungblut2016pathways}.
Researchers have thus been able to perform atomistic simulations in which solid critical nuclei containing hundreds of atoms have been observed to form from the melt~\cite{ten1996numerical,ten1999homogeneous,moroni2005interplay,lechner2011role,prestipino2012systematic}.

When analyzing atomistic simulations of nucleation,  
macroscopic models that provide analytical expressions for the nucleation free energy profile $G(n)$ can be extremely useful
as they can be used to extrapolate the $G(n)$ obtained for the sub-critical nuclei which form in atomistic simulations so as to obtain the free energies of larger nuclei.  This is often essential when predicting $G^{\star}$,
and when interpolating between results obtained at different temperatures.
One simple model which is commonly used to 
rationalize nucleation is classical 
nucleation theory (CNT).  This theory
assumes that $G(n)$ can be expressed as the sum of a bulk 
and a surface term, i.e.
\begin{equation}
    G(n)=\mu n + \gamma \Omega v^{\frac{2}{3}} n^{\frac{2}{3}},
    \label{eq:cnt}
\end{equation}
Here $\mu$ is the chemical potential difference 
between the stable and the metastable phases,
$\gamma$ is the area specific interfacial free energy,
$v$ is the molar volume of the 
bulk stable phase,
and $\Omega$ is a geometrical constant (e.g. $(36 \pi)^{\frac{1}{3}}$ for a spherical nucleus).

CNT has been shown to give rise to significant systematic errors for all sizes of the nucleus~\cite{prestipino2012systematic}.
A large part of problem is that the number of atoms $n$ in the cluster is not a well-defined quantity~\cite{moroni2005interplay,lechner2011role,prestipino2012systematic,cheng2015solid}.
Different definitions of $n$ inevitably affect the calculated free energy profile $G(n)$,
which forces one to ask which definition of $n$ is most appropriate within the CNT framework.
Another part of the problem is that
the CNT expression in Eqn.~\eqref{eq:cnt} is  generally used as a fitting model for $G(n)$,
because
the various physical quantities that enter this expression - the chemical potential $\mu$ and the interfacial free energy $\gamma$ - cannot be calculated directly from homogeneous nucleation simulations.
As a consequence,
when the results from simulations deviate from the predictions of CNT,
it is not clear which assumption within the model is the main culprit.  It is thus difficult
to add additional correction terms to the model to compensate for the missing elements.

An alternative method for probing nucleation involves investigating planar interfaces,
which resemble the surfaces of a large nucleus.
From the simulations of planar interfaces,
the chemical potential as well as the interfacial free energies for different crystallographic directions
can be evaluated~\cite{hoyt2001method,davidchack2000direct,davidchack2003direct,angioletti2010solid,pedersen2013computing,cheng2015solid}.
Various techniques have been developed for computing the interfacial free energy or the stiffness of the planar interfaces at the coexistence temperature~\cite{hoyt2001method,davidchack2000direct,davidchack2003direct,angioletti2010solid,pedersen2013computing}.  
Using the values of $\gamma$ obtained from such simulations within the framework of CNT is problematic though as nucleation occurs in undercooled conditions.
Recently, however, the metadynamics method has been extended so that it also works in out-of-equilibrium conditions away from the melting point~\cite{cheng2015solid}.
This method thus opens the door to a systematic exploration into how the quantities
obtained from simulations of planar interfaces can be used to predict the free energy associated with a three-dimensional nucleus under realistic conditions.

In the present study, 
we apply our recently introduced thermodynamic framework 
for defining the nucleation free energy profile $G(n)$ in an unambiguous way~\cite{cheng2016bridging}.
We then fit the computed $G(n)$ to the expressions that come from macroscopic nucleation models,
using the specific interfacial free energy $\gamma$ and the chemical potential difference $\mu$ as fitting parameters.
We also show that a higher-order correction -- generally referred to as the Tolman term -- naturally emerges from our formalism. 
Finally, in the last part of the paper, we compare the predicted values for $\gamma$ and $\mu$ that we obtain from simulations of three-dimensional nuclei
against those obtained from the simulations of planar interfaces.

\section{Theory}

\subsection{Gibbs dividing surface in atomistic systems}

In order to link the atomistic descriptions and the macroscopic representations of a solid-liquid system,
we define a Gibbs dividing surface using an arbitrarily-chosen extensive quantity $\Phi$ 
that characterises the state of all the atoms within the system.  This extensive quantity might be the volume $V$, the entropy $S$,~\cite{piaggi2016enhancing}
or a global collective variable (CV) $\Phi=\sum_i \phi(i)$ that is constructed by summing the
order parameter values $\phi(i)$ for each of the atoms in the microstate~\cite{cheng2015solid,cheng2016bridging}.
We define the position of the $\Phi$-based Gibbs dividing surface by requiring that there be zero surface excess for the extensive quantity $\Phi$.
In other words, 
the real system and a reference system in which the presence of the interface has no effect on the properties of the two bulk phases are setup so they have the same values for the extensive quantity~\cite{gibbs1928collected,tolman1948consideration}.
Thermal fluctuations ensure that
the instantaneous value of the extensive quantity $\Phi$ 
for this reference system is not fixed, but 
follows a distribution 
$\rho_{\text{ref}}\cond{\Phi}{\ns(\Phi),N-\ns(\Phi)}$ with a 
finite width that can be calculated by performing 
simulations of the two bulk phases~\cite{yukalov2011equilibration,cheng2016bridging}.
To have zero excess for $\Phi$ the real system and the reference system, both of which have $\ns(\Phi)$ solid atoms and $N-\ns(\Phi)$ liquid atoms, should have the same distribution for $\Phi$~\cite{cheng2016bridging}, i.e.
\begin{equation}
    \rhosl\cond{\Phi}{\ns(\Phi),N-\ns(\Phi)} \equiv \rho_{\text{ref}}\cond{\Phi}{\ns(\Phi),N-\ns(\Phi)}.
    \label{eq:gibbsprob}
\end{equation}

In Ref.~\cite{cheng2016bridging},
we argued that Eqn.~\eqref{eq:gibbsprob} can be used
to link a free energy profile $\tilde{G}(\Phi)$, which is expressed as a function of an extensive quantity $\Phi$ for the whole system, with the nucleation free energy for a single solid cluster which is expressed as a function of $\ns(\Phi)$.  
This connection can be made under the  reasonable assumption that the probability of observing a single large sub-critical cluster inside an undercooled bulk liquid can be expressed as $N_s
\exp(-\beta G(\ns(\Phi)))$ when $\ncut \le \ns(\Phi) \le n^{\star}$.  In these expressions $N_s$ is the number of nucleation sites, which in this work we take to be simply the number of atoms, while $G(\ns(\Phi))$ is the free energy excess associated
with the solid cluster. 
Using the law of total probability,
the probability 
distribution for $\Phi$ in such systems
follows
\begin{multline}
        e^{-\beta\tilde{G}(\Phi)}=  \\
        \int_{\ncut}^{n^{\star}} \!\!\!\D \ns(\Phi)
       \rhosl\cond{\Phi}{\ns(\Phi),N-\ns(\Phi)} N_s e^{-\beta G(\ns(\Phi))},
   \label{eq:gphigns}
\end{multline}
This expression is valid for 
values of $\Phi$ that satisfy $\rhosl\cond{\Phi}{\ns(\Phi),N-\ns(\Phi)} \approx 0$ for all $\ns(\Phi) < \ncut$. 
To be clear $\tilde{G}(\Phi)$ can be directly computed from atomistic simulations of nucleation.
Values of $G(\ns(\Phi))$ for $\ns(\Phi) > \ncut$ can then be found by inverting Eqn.\eqref{eq:gphigns} numerically~\cite{cheng2016bridging}. 

\subsection{A general formulation of classical nucleation theory}

With a $\Phi$-based dividing surface,
the free energy of the solid-liquid system relative to that of a bulk liquid can be naturally decomposed into a bulk and a surface term:
\begin{equation}
     G(\ns(\Phi))=\msl \ns(\Phi) + \gsl^{\Phi}A(\ns(\Phi)) ,
     \label{eqn:g}
\end{equation}
Here $\msl=\ms-\ml$ is the difference between the per-atom chemical potentials of the solid and liquid phases.  The
free energy excess associated with the $\Phi$-based dividing surface is then the product of a specific energy term $\gsl^{\Phi}$ and an extensive area term $A(\ns(\Phi))$.

For planar interfaces, the surface area $A(\ns(\Phi))$ in Eqn.~\eqref{eqn:g} is fixed by the boundary conditions so the specific planar interfacial free energy $\gsl^{\Phi} = \gp^{\Phi}$ is
a constant that depends on the crystallographic direction of the planar interface.
Notice that if the extensive quantity $\Theta$ is used to define the dividing surface instead of $\Phi$ the composition of the reference system changes.  This is important as a change in the composition of the reference state will affect the value obtained for the interfacial free energy.  We quantify this change in composition using:
\begin{equation}
    \delta \ns^{\Phi,\Theta}=(\ns(\Theta)-\ns(\Phi))/A
    \label{eq:dns}
\end{equation}
and further note that this quantity should be constant as replacing $\Phi$ with $\Theta$ shifts the location of the dividing surface by a fixed amount~\cite{cheng2015solid}.
Furthermore, when $\Phi$ is replaced by $\Theta$
the resulting change in the planar interfacial free energy
\begin{equation}
     \Delta \gp^{\Phi,\Theta}=\gp^{\Theta} - \gp^{\Phi} = -\msl \delta \ns^{\Phi,\Theta},
     \label{eq:gthetaphi}
\end{equation}
is also a constant, because the total free energy in Eqn.~\eqref{eqn:g}
 should be unaffected by the change in the extensive quantity~\cite{cheng2015solid}. 

For the curved interfaces around a three-dimensional nucleus,
the surface area $A(\ns(\Phi))$ is not
equally well-defined.
Classical nucleation theory assumes that the nucleus has the same density as the bulk solid, 
which is what ensures that $A= \Omega \vs^{\frac{2}{3}} \ns^{\frac{2}{3}}$.
Critically, however, this bulk density assumption is only valid when the equimolar dividing surface is used to define the reference state,  
which implies that only quantities determined using a dividing surface with no excess volume should be inserted into Eqn.~\eqref{eqn:g}.
This equation should thus read:
\begin{equation}
 G(\ns(V))=\msl \ns(V) + \gsl^{V} \Omega \vs^{\frac{2}{3}} \ns^{\frac{2}{3}}(V).
 \label{eq:cntv}
\end{equation}

Even when the reference state is defined in a manner consistent with this assumption about the bulk density of the nucleus, however, one still   
 has to include a term that incorporates the surface excess free-energies' dependence on curvature into  $\gsl^{V}$~\cite{tolman1948consideration,block2010curvature}.
This dependence of $\gsl^{V}$ 
on the effective radius of the nucleus  $R = (3/4\pi)^{\frac{1}{3}} \vs^{\frac{1}{3}} \ns^{\frac{1}{3}}(V)$ can be written as 

\begin{equation}
\gsl^{V}(R)=\gp^{V}(1-2\delta/R+\mathcal{O}(1/R^2)).
 \label{eq:tolmanlength}
\end{equation}
In this expression $\delta$ is the Tolman length - a quantity that measures the difference between the locations of the equimolar dividing surface and the surface of tension at the planar limit.  $\gp^{V}$, meanwhile, is the specific interfacial free energy associated with the equimolar surface in the planar limit.
Taking only the leading terms in Eqn.~\eqref{eq:tolmanlength},
Eqn.~\eqref{eq:cntv} can be rewritten as
\begin{equation}
 G(\ns(V))\eqsim \msl \ns(V) + \gp^{V} \Omega \vs^{\frac{2}{3}} \ns^{\frac{2}{3}}(V)
 (1- \epsilon \ns^{-\frac{1}{3}}(V)),
 \label{eq:cntv2}
\end{equation}
where $\epsilon = (32\pi/3)^{\frac{1}{3}}\vs^{-\frac{1}{3}}\delta$ is a constant.

At the atomic scale, due to the magnitude of fluctuations, it is difficult to distinguish between the different phases based only on the difference in molar volume. Hence, it is often not convenient to use this quantity to determine the number of solid particles in a simulation.
One can use a different extensive quantity to determine the location of the diving surface as it is possible to convert the free-energy profile obtained using one dividing surface to 
the result that would have been obtained for a different dividing surface using Eq.~\eqref{eq:dns}~\cite{cheng2015solid}. It would be useful, however,
to derive a CNT expression 
based on an arbitrary choice
of reference thermodynamic variable $\Phi$.
To derive such an expression,
we first assumed that the difference between the location of the $\Phi$-based dividing surface and the location of the equimolar dividing surface is much smaller than the effective radius of the nucleus $R$, as this ensures
that the difference between $\ns(\Phi)$ and $\ns(V)$
can be approximated using
\begin{equation}
    \ns(V)-\ns(\Phi) \eqsim \delta \ns^{\Phi,V} \Omega \vs^{\frac{2}{3}} \ns^{\frac{2}{3}}(V),
\end{equation}
where $\delta \ns^{\Phi,V}$ is the constant that appears in Eqn.~\eqref{eq:dns} and ~\eqref{eq:gthetaphi} and where $\Omega \vs^{\frac{2}{3}} \ns^{\frac{2}{3}}(V)$ is a measure of the surface area of the nucleus.
We can substitute $\ns(V)$ for $\ns(\Phi)$ in Eqn.~\eqref{eq:cntv2} using the above approximation,
and can also use the relation $\gp^{V} - \gp^{\Phi} = -\msl \delta \ns^{\Phi,V}$.
After dropping the higher order correction terms, we obtain
\begin{equation}
    G(\ns(\Phi))
    \eqsim \msl \ns(\Phi) + \gp^{\Phi}\Omega \vs^{\frac{2}{3}}  \ns^{\frac{2}{3}}(\Phi)
    (1 +\zeta \ns^{-\frac{1}{3}}(\Phi))
     ,
    \label{eq:tolphi}
\end{equation}
where $\zeta=\frac{2}{3}\delta \ns^{\Phi,V}  \Omega \vs^{\frac{2}{3}} - \epsilon $ is a constant.
This $\zeta \ns^{-\frac{1}{3}}(\Phi)$ term in Eqn.~\eqref{eq:tolphi}, which has the same mathematical form as the Tolman correction,
has two components:
the $\delta \ns^{\Phi,V}$ term stems from the difference between the $\Phi$-based dividing surface and the equimolar surface,
while the other $\epsilon$ term is determined from the Tolman length and has a value that is independent of the choice of the dividing surface.
Hereafter, we will use CNT($\Phi$)+Tol to denote the expression in Eqn.~\eqref{eq:tolphi} in order to highlight its dependence on the chosen extensive quantity. An awareness of this dependence on the choice of $\Phi$ is crucial when comparing 
studies performed with different 
protocols. Eqs.~\eqref{eq:cntv2} and \eqref{eq:tolphi} are both valid 
expressions for the free energy of a nucleus that are consistent with 
Tolman-corrected CNT.  These two expressions differ, however, when they come to defining the
size of the nucleus and the value and interpretation of the planar-interface
excess free energy and the coefficient for 
the finite-size correction.

\section{Simulation methods}

We simulated the processes of homogeneous nucleation for 
a simple but realistic Lennard-Jones system~\cite{davidchack2003direct,angi+10prb,benjamin2014crystal}.
The NPT ensemble was employed throughout with the  Nose-Hoover thermostat and isotropic barostat.
The time step was set equal to 0.004 Lennard-Jones time units and a supercell containing 23328 atoms was used throughout.  
\begin{figure}[btp]
\includegraphics[width=0.5\textwidth]{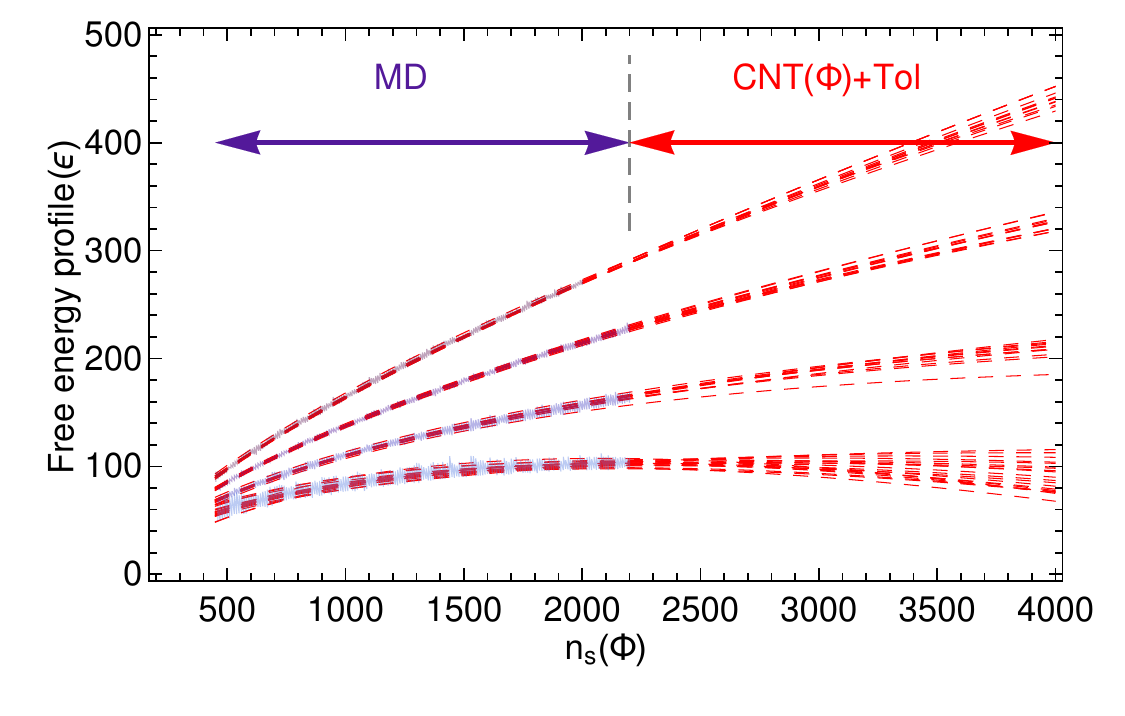}
\caption{
The four sets of curves from  bottom to top correspond to the free energy profiles of $\ns(\Phi)$ at temperatures of $0.56$, $0.58$, $0.60$ and $0.6185$, respectively.
Each thin blue curve is computed from a biased molecular dynamics simulation,
and each thin red line corresponds to the best CNT($\Phi$)+Tol fit of one blue curve.
}
\label{fig:fes}
\end{figure}

Each independent simulation at each temperature was run for approximately $6 \times 10^6$ steps.
To accelerate the sampling so as to obtain reversible 
formation of a solid nucleus in a viable 
amount of simulation time,
we performed biased sampling using the well-tempered metadynamics 
protocol with adaptive Gaussians ~\cite{barducci2008well,branduardi2012metadynamics},
and the collective variable $\Phi=\sum_i S(\kappa(i))$ that was employed in Ref.~\cite{cheng2015solid}.
In essence, $S(\kappa(i))$ is a local structural fingerprint for atom $i$,
which takes a value of $1$ for a perfect \emph{fcc} crystal and $0$ for a homogeneous liquid.
Fast implementation of this complex 
simulation setup was made possible by the flexibility of the PLUMED code~\cite{tribello2014plumed} in combination with LAMMPS~\cite{plim95jcp}.
See the Supplemental Material for sample input files~\cite{SI}.
In addition to the simulations of homogeneous nucleation,
we also simulated planar interfaces along the $\avg{100}$, $\avg{111}$, and $\avg{110}$
crystallographic directions of the \emph{fcc} lattice.
For the planar interface simulations, the setups are identical to the ones described in Ref.~\cite{cheng2015solid}.

\section{Results and Discussions}
At each of the temperatures $0.56$, $0.58$, $0.60$ and $0.6185$, a total of 12 independent metadynamics runs were performed.
For each simulation run we first computed the free energy profile $\tilde{G}(\Phi)$ with
respect to the extensive quantity $\Phi$ of the system.  We then extracted the nucleation free energy for a single solid cluster as a function of $\ns(\Phi)$ using the framework
introduced above, which is thoroughly described in Ref.~\cite{cheng2016bridging}.
Each free energy profile $G(\ns(\Phi))$ is plotted as a thin blue curve in Figure~\ref{fig:fes}.
For each $G(\ns(\Phi))$, we then performed a CNT($\Phi$)+Tol fit using Eqn.~\eqref{eq:tolphi} with $\msl$, 
$\gp^{\Phi}\Omega$, and $\zeta$ as fitting parameters.  Each of these fitted curves is shown in red in Figure~\ref{fig:fes}.

\begin{figure}

\includegraphics[width=0.5\textwidth]{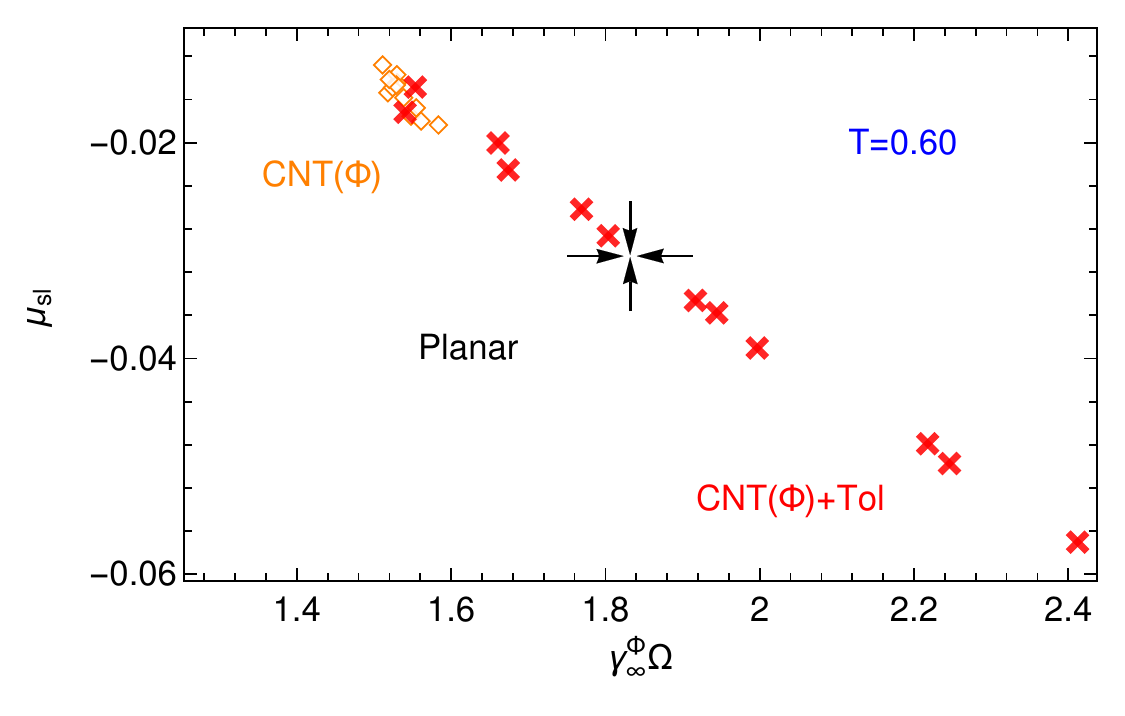}
\caption{
Each symbol in red or orange represents one pair of fitting parameters ($\gp^{\Phi} \Omega$ and  $\msl$) that were obtained by fitting one of the free energy curves shown in figure \ref{fig:fes} that were themselves extracted from one of the independent simulations that were performed at $T=0.60$.
The black arrows indicate the results obtained from simulations of planar interfaces.
All quantities are
expressed in Lennard-Jones units.
}
\label{fig:fits}
\end{figure}

\begin{figure}

\includegraphics[width=0.5\textwidth]{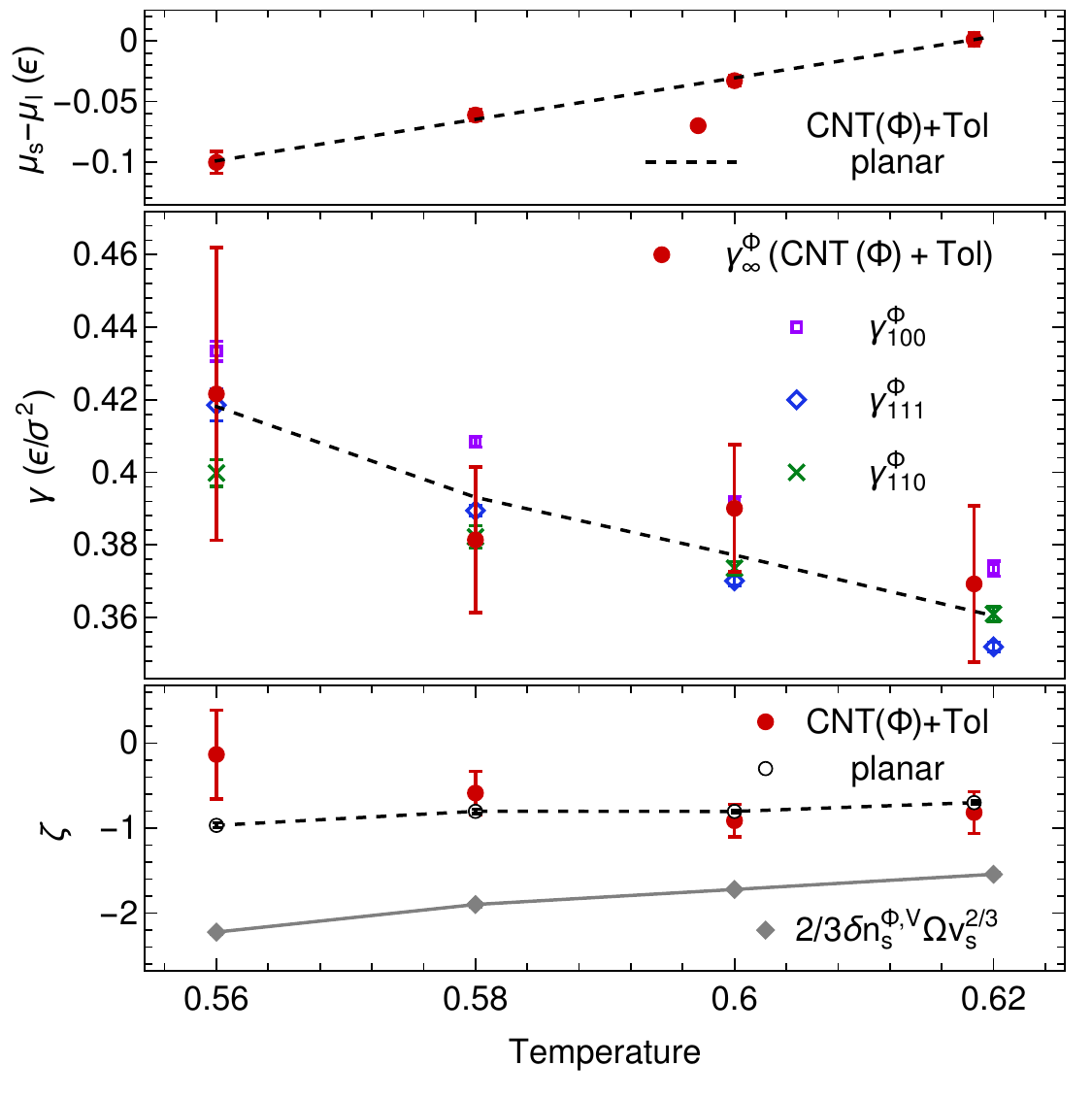}
\caption{
In the top panel, 
the red symbols indicate $\msl$ values computed from homogeneous nucleation simulations using the CNT($\Phi$)+Tol fit,
the dashed line $\msl=1.714(T-0.6178)$ describes the fitted values of $\msl$ that was extracted from planar interface calculations using a large simulation box with 20736 atoms.
In the middle panel, the purple, blue and green symbols indicate the values of $\gp^{\Phi}$ for the planar interfaces that are perpendicular to three different lattice directions.  The dashed line indicates the effective $\gp^{\Phi}$ that is obtained when the results for the planar interfaces are averaged over all lattice directions.
The red symbols indicate the values of the estimates for the fitting parameters $\gp^{\Phi} \Omega$ that are obtained from the CNT($\Phi$)+Tol model divided by the estimated $\Omega$ at each temperature.
In the bottom panel, the red symbols indicate the Tolman correction constant $\zeta$ in the CNT($\Phi$)+Tol fits, while
the black symbols indicate the estimate for $\zeta$ that is obtained when the planar interface results for $\msl$ and $\gp^{\Phi}$ are used in Eqn.~\eqref{eq:tolphi}.
The gray symbols indicate the equimolar-surface correction to $\gamma_\infty^\Phi$ from the planar limit, $\frac{2}{3}\delta \ns^{\Phi,V}\Omega \vs^{\frac{2}{3}}$. 
Statistical uncertainties are indicated throughout using error bars.
}
\label{fig:se}
\end{figure}

\begin{table*}
\caption{A comparison of the predictions for $\msl$ and $\gp^{\Phi}$ from different models. For each number, the value in the bracket indicates the statistical uncertainty in the last digit.}
\label{tab:gm}
\small
    \begin{tabular}{ p{2cm} | c c c c | c c c c }
   \hline\hline
  & \multicolumn{4}{ c| }{\msl} & \multicolumn{4}{ c }{$\gp^{\Phi}$} \\ 
  model & 0.56 & 0.58 & 0.60 & 0.6185 & 0.56 & 0.58 & 0.60 & 0.6185\\\hline   
 CNT($\Phi$)+Tol & -0.100(9) & -0.061(5) & -0.033(4) & 0.001(5) &
  0.42(4) & 0.38(2) & 0.39(2) & 0.37(2)\\ 
 Planar & -0.0991(2) & -0.0648(2) & -0.0305(2) & 0.0012(2) & 
  0.418(4) & 0.393(3) & 0.377(2) & 0.360(2)\\  
 CNT($\Phi$) &  -0.083(2) & -0.049(1) & -0.016(1) & 0.017(1) &
  0.346(5) & 0.330(2) & 0.317(1) & 0.302(3)\\ \hline \hline
    \end{tabular}
\end{table*}

The red curves in Figure~\ref{fig:fes} start to diverge at larger sizes even though they almost overlap at small sizes.
In order to understand the origin of this divergence we show the two parameters ($\msl$, $\gp \Omega$) obtained from CNT+Tol($\Phi$)
fits for each free energy profile at $T=0.60$ in Figure \ref{fig:fits}.
The two parameters $\msl$ and $\gp\Omega$ are clustered around a straight regression line in Figure~\ref{fig:fits}, which 
suggests that there is a very large correlation between them. 
The consequence of this correlation is that a tiny change in the segment of data used for fitting can make the two parameters vary collectively along the regression line by a significant amount. 
Thus, small uncertainties in the computed free energy profile for the small sub-critical nuclei that form in simulations can propagate and amplify when these curves are extrapolated to large nuclei. 

We wanted to determine whether or not 
the values obtained for $\msl$ and $\gp\Omega$ from the CNT($\Phi$)+Tol model are physically meaningful.
To this end, we also show results obtained from simulations of the planar interfaces in Figure \ref{fig:fits}. 
The value of $\msl$ for an interface perpendicular to any lattice direction can be computed by performing biased simulations of that interface~\cite{pedersen2013computing,cheng2015solid}.
We computed these planar interfacial free energies  in our previous work~\cite{cheng2015solid} and used a deterministic framework to locate the Gibbs dividing surface.
In the Supplementary Material,
we show that the values of the interfacial free energies do not change significantly when the probabilistic framework in  Eqn.~\eqref{eq:gibbsprob} is applied so
 for this reason we chose to adopt the values for $\gamma^{\Phi}$ and $\gamma^V$ reported in Ref.~\cite{cheng2015solid}.
To estimate $\gp^{\Phi}\Omega$,
we first obtained the values of $\gamma_{100}^{\Phi}$, $\gamma_{111}^{\Phi}$ and $\gamma_{110}^{\Phi}$ for planar interfaces with normal vectors parallel to the specified lattice directions.
We then used a common assumption; namely, that for this Lennard-Jones system the interfacial free energy surface $\gamma^{\Phi} (\vec{n})$ can be expanded using cubic harmonics to the third order~\cite{hoyt+01prl},
whose coefficients can be parametrized using the values of $\gamma_{100}^{\Phi}$, $\gamma_{111}^{\Phi}$ and $\gamma_{110}^{\Phi}$.
As shown in the Supplementary Material,  the shape of the equimolar surface of the nucleus $R(\vec{n})$ can then be reproduced by performing a Wulff construction using the $\gamma^{V}$ values~\cite{adams1994interfacial,SI}.
As the difference between the locations of the $\Phi$-based dividing surface  and  the  equimolar  dividing surface was assumed to be insignificant compared to the radius of the nucleus,
the shape, $R(\vec{n})$, that emerges from the Wulff construction can be considered to be very close to the shape of the $\Phi$-based dividing surface of the nucleus.
From this shape,  
we can, therefore, estimate the geometrical constant $\Omega$,
and an effective value of $\gp^{\Phi}$ for the surface of the whole nucleus 
by computing the surface integral $\iint_R \gamma^{\Phi} (\vec{n}) \,dA / \iint_R \,dA$.
At all the temperatures we considered,
the shapes $R(\vec{n})$ that we obtained were close to spherical.  In fact, the corresponding $\Omega$ values were within 0.5\% of the geometrical constant for a sphere.

Figure \ref{fig:fits} suggests that the values of $\msl$ and $\gp^{\Phi}\Omega$ from the CNT($\Phi$)+Tol fits
are consistent with the planar limit results.  However, there is a large spread in the parameter values because of the strong correlation between these two parameters in the fitting.
Table~\ref{tab:gm} shows all the values of $\msl(T)$ together with the values of $\gp(T)$ that are predicted both from the CNT($\Phi$)+Tol model and the planar limit results and further illustrates the
very good agreement between the two models at all temperatures considered.
We think this good agreement is a strong indication that
the  CNT($\Phi$)+Tol model is a good macroscopic model for the free energies of atomistic nuclei.
Indeed, when a different model is used, the values of the two parameters $\msl$ and $\gp^{\Phi}\Omega$ may not agree with planar limit results.
For example, we also performed a so-called CNT($\Phi$) fit on each computed $G(\ns(\Phi))$,
using a conventional CNT formulation $G(\ns(\Phi))=\msl \ns(\Phi) + \gp^{\Phi} \Omega \vs^{\frac{2}{3}} \ns^{\frac{2}{3}}(\Phi)$~\cite{SI}.
The values of $\msl$ and $\gp^{\Phi}\Omega$ from the CNT($\Phi$) fits are indicated in Figure~\ref{fig:fits} and Table~\ref{tab:gm}.  Once again there are very strong inter-correlations between the parameters.  What is even worse, however, is that the fitted values are no longer consistent with the planar interface results.
This example highlights the perils associated with using the wrong model to fit the free energy profile for nucleation:
although one might obtain a good fit for the data points,
the values of the fitted parameters of the model are physically unrealistic and the model will most likely have limited predictive power in scenarios that are outside the range of the existing data set.

The comparison between the values of $\msl$ and $\gp^{\Phi}\Omega$ from the CNT($\Phi$)+Tol and the planar limit results suggests that
one should just use the results obtained at the planar interface limit in Eqn.~\eqref{eq:tolphi}
so as to have a model that has $\zeta$ as the only fitting parameter.  
This model would have
much greater statistical accuracy as there are no correlations between the fitting parameters.
We performed such fittings and compared the values of the parameter $\zeta$ with the values obtained from the regular CNT($\Phi$)+Tol fits using all three parameters.
The bottom panel of Figure~\ref{fig:se}, shows that this new
approach results in much smaller uncertainties in the values of $\zeta$.
Therefore, performing simulations of planar interfaces as well as simulations of three dimensional nuclei is worthwhile
when studying homogeneous nucleation, 
especially given how computationally inexpensive such simulations are.
We would recommend first computing the values of $\msl$ and $\gp^{\Phi}\Omega$ from simulations of planar interfaces.  This allows one
to determine the leading terms $\msl \ns(\Phi) + \gp^{\Phi} \Omega \vs^{\frac{2}{3}} \ns^{\frac{2}{3}}(\Phi)$ in a CNT-type model.
Comparing these leading terms with the actual nucleation free energy profiles computed from the simulations of three-dimensional nuclei,
then allows one to extract the other higher-order correction terms in the CNT-type model 
with much higher statistical accuracy.

It is worth discussing the higher-order correction term
$\zeta=\frac{2}{3}\delta \ns^{\Phi,V}  \Omega \vs^{\frac{2}{3}} - \epsilon $ in Eqn.~\eqref{eq:tolphi} a little further.
This factor has two distinct components but
we treat it as a single fitting parameter in the present study.
To evaluate the first term separately we can exploit the fact that
the difference between the number of solid-like atoms per area $\delta \ns^{\Phi,V}$ for different definitions of the dividing surface
can be directly evaluated from simulations of planar interfaces as discussed in Ref.~\cite{cheng2015solid}.
We evaluated $\delta \ns^{\Phi,V}$ for planar interfaces with normal vectors parallel to the $\avg{100}$, $\avg{111}$ and $\avg{110}$ lattice
directions at each temperature,
and then approximated $\delta \ns^{\Phi,V}(\vec{n})$ using a cubic harmonic expansion,
and thus estimated its averaged value for a three-dimensional nucleus.
These estimated values for $\frac{2}{3}\delta \ns^{\Phi,V}  \Omega \vs^{\frac{2}{3}}$ are shown
in the bottom panel of Figure~\ref{fig:se}.
The other term $\epsilon=(32\pi/3)^{\frac{1}{3}}\vs^{-\frac{1}{3}}\delta$ that enters $\zeta$ can be determined from the Tolman length $\delta$.  This quantity 
can be evaluated from the pressure tensor at each position along the normal direction of a planar interface.  However, because calculations of this sort are difficult to converge~\cite{lei2005tolman,blokhuis2006thermodynamic,van2009direct}, we chose to subtract the values we obtained for $\frac{2}{3}\delta \ns^{\Phi,V} \Omega \vs^{\frac{2}{3}}$ from the value we obtained for $\zeta$ and to extract the Tolman length from $\epsilon$.
In this way, we predicted the Tolman length to be on the order of $-0.3$ Lennard-Jones length units at the temperatures considered.

Finally,
we want to point out that
the fact that the CNT($\Phi$)+Tol model 
succeeds in
describing small nuclei that contain hundreds of atoms, while being
consistent with simulations performed in the limit of a planar solid-liquid interface 
has theoretical implications.
Even though the actual interface between the solid and the liquid is diffuse and fluctuating,
the concept of the Gibbs dividing surface can be used to convert the atomistic descriptions of the solid-liquid system into a macroscopic representation.  Moreover, finite-size effects are well-captured by simple corrections that
take into account the deviation between 
the chosen dividing surface, the equimolar 
dividing surface and the surface of tension. In this case, these two corrections are of comparable size.

\section{Conclusions}

In summary we have studied, by means of atomic-scale simulations,
 the various different terms that appear in  classical nucleation theory. 
 When analysing our simulations we define a Gibbs dividing surface from the value of a macroscopic order parameter when calculating the dependence of the free energy on the number of particles in the nucleus. Then, by juxtaposing explicit simulations of a three-dimensional nucleus with simulations of a planar solid-liquid interface, performed in equivalent thermodynamic conditions, we identify the effects that lead the results on small nuclei to deviate from the predictions of 
classical nucleation theory. We find that deviations occur because curvature-dependent corrections
to the planar-interface surface
energy (the so-called Tolman term) are required.   These genuine departures from CNT should not be confused with the deviations that occur when an arbitrary order parameter is used to identify the nucleus, however.  Such corrections are required simply because we require zero surface excess for the arbitrary extensive quantity $\Phi$ when calculating the surface excess free energy.  As $\Phi$ is not the volume the value we obtain for this excess free energy differs from the value that would have been obtained had we found the true equimolar dividing surface.  The 
leading order correction for this effect has the same functional form as the Tolman term.  It should not, however, be considered as a genuine departure from CNT as it is simply an artifact in the analysis.

In the final parts of the manuscript we discussed the statistical efficiency of different approaches for finding the CNT parameters. We showed that the surface and bulk parameters in the CNT model are strongly correlated, which makes the fitted values of these parameters extraordinarily sensitive to small deviations in the computed free energy surfaces that are used to fit them.  We thus concluded that fitting these terms using a free energy surface obtained from a simulation of a 3D nucleus is not optimal and that also using information from simulations of planar-interfaces is thus beneficial.   
In fact, given that planar-interface 
models converge faster with respect to both size and 
simulation time, and that such simulations can be more easily analyzed in 
terms of different order parameters, we suggest that they should always be used in a preliminary phase.  Explicit simulations of 3D nucleation can then be used to identify genuine finite-size effects.
We hope that our careful analysis will help resolve some
of the ambiguities in atomic-scale studies of both homogeneous and heterogeneous nucleation, and that this work will lay the foundations
for the modelling of more complex materials and for the study of different kinds of phase transitions.

\end{document}